\documentclass[showpacs,preprintnumbers,aps,prl,letterpaper,superscriptaddress,nofootinbib,tightenlines,floats,floatfix]{revtex4}
\usepackage{graphicx}
\usepackage{bm}
\usepackage{latexsym}
\usepackage{amsmath,amssymb}
\usepackage{amsfonts}
\usepackage[draft=false]{hyperref}
\usepackage[latin1]{inputenc}
\usepackage{amsthm}
\usepackage{mathrsfs}
\usepackage{ifpdf}
\usepackage{color}
\usepackage{epstopdf}
\usepackage[skip=2pt,font=scriptsize]{caption}
\usepackage{natbib}
\usepackage{blindtext}
\allowdisplaybreaks

\begin{document}
\title{Gravitational perturbations in the Newman-Penrose formalism: Applications to wormholes}

\author{Juan Carlos Del \'Aguila}
\email{jdelaguila@fis.cinvestav.mx}
\affiliation{Departamento de F\'isica, Centro de Investigaci\'on y de
  Estudios Avanzados del IPN, A.P. 14-740, 07000 Ciudad de M\'exico, M\'exico.}
\author{Tonatiuh Matos}
\email{tmatos@fis.cinvestav.mx}
 \altaffiliation{Part of the Instituto Avanzado de Cosmolog\'ia (IAC)
  collaboration http://www.iac.edu.mx/}
\affiliation{Departamento de F\'isica, Centro de Investigaci\'on y de
  Estudios Avanzados del IPN, A.P. 14-740, 07000 Ciudad de M\'exico,
  M\'exico.}

\begin{abstract}
In this work we study the problem of linear stability of gravitational perturbations in stationary and spherically symmetric wormholes. For this purpose, we employ the Newman-Penrose formalism which is well-suited for treating gravitational radiation in General Relativity, as well as the geometrical aspect of this theory. With this method we obtain a ``master equation'' that describes the behavior of gravitational perturbations that are of odd-parity in the Regge-Wheeler gauge. This equation is later applied to a specific class of Morris-Thorne wormholes and also to the metric of an asymptotically flat scalar field wormhole. The analysis of the equations that these space-times yield reveals that there are no unstable vibrational modes generated by the type of perturbations here studied.
\end{abstract}

\date{Received: date / Accepted: date}

\maketitle

\date{\today}


\maketitle

\section{I. Introduction}

Ever since the concept of ``wormhole'' was first introduced in the literature by Misner and Wheeler in 1957 \cite{wheeler}, there has been general interest in the fascinating physical and geometrical properties that the non-trivial topology of these objects could possess. Previously, Einstein and Rosen had already proposed an interpretation of the Schwarzschild space-time consisting of two identical ``sheets'' connected through a ``bridge'', this is nowadays known as an Einstein-Rosen bridge \cite{ER}. While the Scwarzschild metric can indeed admit the interpretation of wormhole, Fuller and Wheeler would later prove that it is not a traversable one since, in order for a signal to cross it, causality must be violated \cite{fuller}. Nevertheless, these ideas can be considered as the precursors of the modern version of a wormhole, an hypothetical compact object that would allow communication between distant regions of the same universe, or even between two different universes. Several years later, Morris and Thorne established the characteristics that a traversable wormhole must feature, along with a general metric that should describe them \cite{mtwh}. The metric was assumed to be stationary and spherically symmetric. In their work they arrived to the unfortunate conclusion that a traversable wormhole requires of ``exotic'' matter, this is, matter that violates the energy conditions, as its gravitational source. Until now there is no observational evidence on the existence of wormholes, and thus, they have remained within the speculative realm of the theory of General Relativity. Moreover, there are additional concerns regarding the realistic existence of such objects in the universe, one of them is that of their stability.

It can be argued that, in order for a stellar object to be of any astrophysical interest, it has to stable. Otherwise, a slight perturbation or deviation from its initial state would result in the collapse of the space-time itself. Remarkable papers that solve this problem, at least in a linear theory and using analytical methods, are well-known. The stability of the Schwarzschild metric was first studied by Regge and Wheeler by adding a perturbation term $h_{\mu\nu}$ to the background metric and keeping terms up to first order of the perturbation \cite{rw}. Based on their results, later works then confirmed that said space-time is indeed stable against small perturbations \cite{vish,Zerilli}. With a less straightforward approach, Teukolsky managed to find a ``master equation'' that describes gravitational, electromagnetic, and neutrino field perturbations of a spinning black hole \cite{teukolsky1}. Analyzing then the mentioned equation with numerical techniques, the stability of the first modes of vibration of the Kerr black hole was concluded \cite{teukolsky2}. From a practical point of view, this result was of great relevance to the existence and possible observation of a realistic black hole. However, from a theoretical standpoint, maybe the most interesting aspect of this series of papers is that of the derivation of the master equation. For this purpose, Teukoslky exploited the full potential of the Newman-Penrose formalism \cite{np} and the underlying geometric properties of the Kerr metric, in particular, the fact that it is of type D in the algebraic classification of space-times. As impressive as the master equation is, unfortunately, it is only valid for vacuum space-times of type D. This rules out the possibility of applying it to wormholes. Another work in which the Newman-Penrose formalism is applied in the context of perturbation theory can be found in \cite{hamiltonNP}, where the main focus is centered on self-similar black holes.

It is clear that the problem of stability for black holes has been thoroughly studied. Over the years, these developments have contributed to the physical relevance of this outstanding prediction of General Relativity. On the other hand, wormholes are still only theoretical entities. They are commonly, but not uniquely, proposed as stationary and spherically symmetric space-times supported by a phantom scalar field, i.e., a scalar field whose kinetic energy has a reversed sign (sometimes referred too as ghost scalar field). Maybe the most simple model of such a wormhole is that of Ellis and Bronnikov \cite{ellis,Bronnikov}. In recent years, many works have been written about the question of the stability of these type of wormholes and a handful of them report that they are generally unstable. Thus adding another problematic issue to their set of particular properties. Shinkai and Hayward showed numerically that the Ellis wormhole would collapse to a black hole or form an inflationary universe, depending on the type of matter that causes the perturbation \cite{shinkai}. Analytically, it is also firmly established that the radial monopole mode of phantom wormholes unavoidably leads to instabilities \cite{phantom,phantom2,phantom3}. Particularly in \cite{phantom3}, the case of the Ellis wormhole was studied exhaustively. These results were later extended to include a phantom scalar field with a self-interacting potential \cite{phantom1,phantom4}. In both of these references specific examples of scalar field wormholes consisting either of asymptotically flat, or (A)dS ends (or a mixture of them), were also considered. On the contrary, a stable wormhole was reported in \cite{est3}. Curiously enough, its metric is that of Ellis, however, instead of being supported by the usual phantom scalar field, its gravitational source is constituted by two elements: a radial electric field (a magnetic field is also possible) without sources and a perfect fluid with negative density. Other stable models are obtained through the use of thin shells of matter \cite{est2}. Alternatively, it has been proposed as a conjecture that rotation in the metric of wormholes may induce stability against gravitational perturbations. Several wormholes models have been generalized to include said rotation, see \cite{ring,Matos} for just a few examples. Despite this, and mainly due to the complexity on the geometry of the space-time, the treatment of the perturbation equations can become rather cumbersome. 

In this context, our intention in this paper is to develop a basic framework for treating linear gravitational perturbations using the Newman-Penrose formalism. This will benefit us with some previously used tools and results that have been found over the years for the problem of gravitational radiation in General Relativity. Additionally, in this first approach, we will specialize to odd-parity perturbations, named so by Regge and Wheeler \cite{rw}, in spherically symmetric space-times. The results we obtain will later on be applied to wormhole metrics, some of them supported by phantom scalar fields. Though not treated yet in this paper, we believe that through this scheme, a description of the perturbations in geometrically more complicated space-times could be facilitated. The work is organized as follows. In section II we will study the general problem of gravitational perturbations within the tetrad formalism. This treatment will be particularized to stationary and spherically symmetric space-times in section III, here, the master equation will also be presented.  In section IV the meaning of physical regularity that must be imposed to the mentioned perturbations will be discussed. Examples of the application of our master equation will be given in sections V and VI, first on the Morris-Thorne wormholes, and then on a specific phantom scalar field wormhole. Finally, conclusions and final comments will be given. We have also included two appendixes containing mostly laborious calculations that were carried out throughout the paper.  

\section{II. Gravitational Perturbations in the Tetrad Formalism}

We shall follow the notation utilized by Newman and Penrose in their seminal paper \cite{np} and hereafter may refer to this work as the NP paper. In this formalism a null tetrad $(l^\mu,n^\mu,m^\mu,\bar{m}^\mu)$ is introduced into every point of a four-dimensional pseudo-Riemannian manifold of signature $(+1,-1,-1,-1)$ and metric $g_{\mu\nu}$. The vectors $l^\mu$ and $n^\mu$ are real, while $m^\mu$ and $\bar{m}^\mu$ are complex. In this paper we will use a bar over any given quantity to denote its complex conjugate. The vectors of the tetrad must also satisfy the orthogonal property $l^\mu n_\mu=-m^\mu\bar{m}_\mu=1$, with the rest of the vector combinations being zero. The space-time metric can then be expressed as

\begin{equation}
g_{\mu\nu}=l_\mu n_\nu+n_\mu l_\nu-m_\mu\bar{m}_\nu-\bar{m}_\mu m_\nu.
\label{guv}
\end{equation}

This relation can be rewritten in a more compact way as $g_{\mu\nu}=z_{m\mu}z_{n\nu}\gamma^{mn}$ if one conveniently defines\footnote{We will use Greek indices $(\mu,\nu=0,1,2,3)$ to denote tensor indices and lower-case Latin indices $(a,b,m,n=0,1,2,3)$ to denote tetrad indices.} 

\begin{eqnarray}
z_{m\mu}&=&(l_\mu,n_\mu,m_\mu,\bar{m}_\mu), \nonumber\\
z_m^{\ \mu}&=&(l^\mu,n^\mu,m^\mu,\bar{m}^\mu), \nonumber\\
\gamma_{mn}=\gamma^{mn}&=&
\begin{bmatrix}
	0 & 1 & 0 & 0 \\
	1 & 0 & 0 & 0 \\
	0 & 0 & 0 & -1 \\
	0 & 0 & -1 & 0 \\ 	
\end{bmatrix},
\label{zmu}
\end{eqnarray}  

where $\gamma^{mp}\gamma_{pn}=\delta^m_{\ n}$. Using (\ref{zmu}) we can also write the orthogonality properties simply as $z_m^{\ \mu}z_{n\mu}=\gamma_{mn}$. The metric $\gamma$ will be used to raise or lower tetrad indices. Newman and Penrose define 12 complex spin coefficients that depend on linear combinations of the quantities $\mathcal{Z}_{mnp}=z_m^{\ \mu}z_n^{\ \nu}\nabla_\mu z_{p\nu}$, which are anti-symmetrical in their last two indices\footnote{In the case of $\gamma$ and $\mathcal{Z}_{mnp}$ our notation will slightly vary from that of the NP paper. They use $\eta$ instead of $\gamma$ for the tetrad metric, and $\gamma_{pnm}$ instead of $\mathcal{Z}_{mnp}$. Note that the order of the indices for these last quantities is also different.}, i.e., $\mathcal{Z}_{mnp}=-\mathcal{Z}_{mpn}$. Additionally, four differential operators are introduced

\begin{equation}
D=l^\mu\nabla_\mu, \hspace{3mm} \Delta=n^\mu\nabla_\mu, \hspace{3mm} \delta=m^\mu\nabla_\mu, \hspace{3mm} \delta^*=\bar{m}^\mu\nabla_\mu, 
\label{Dm}
\end{equation}

or more compactly $D_m=z_m^{\ \mu}\nabla_\mu$ with $D_m=(D,\Delta,\delta,\delta^*)$. Using the 12 spin coefficients, along with the operators (\ref{Dm}), Newman and Penrose obtained a set of numerous equations that are the equivalent of the Bianchi identities and the components of the Ricci and Weyl tensors in tetrad form, this is now known as the Newman-Penrose formalism. Since the Einstein field equations make use of the curvature tensors yielded by a given space-time metric, one can discuss any problem in General Relativity (at least its geometrical aspects) within this formalism.

Here we will develop a general framework for perturbation theory using the Newman-Penrose formalism. The scheme we follow is the typical one for linear gravitational perturbations, that is, we add a perturbation term $h_{\mu\nu}$ to a certain background metric $g_{\mu\nu}$, and then compute the components of the Ricci tensor keeping terms up to first order of the perturbation. The perturbation term is assumed to be small compared to its background counterpart. In this formalism, the perturbation term of the metric will be represented by a perturbation in the null tetrad, for example, $l_\mu=\widetilde{l}_\mu+\hat{l}_\mu$, where we will establish the convention that a tilde denotes any given background quantity and the hat denotes the perturbation term of said quantity. To proceed, we expand the perturbation terms of the tetrad in the basis of the background tetrad, hence, 

\begin{eqnarray}
z_m^{\ \mu}&=&\widetilde{z}_m^{\ \ \mu}+\hat{z}_m^{\ \ \mu}=\widetilde{z}_m^{\ \ \mu}+\hat{\Sigma}_m^{\ \ n}\widetilde{z}_n^{\ \mu}, \nonumber\\
z_{m\mu}&=&\widetilde{z}_{m\mu}+\hat{z}_{m\mu}=\widetilde{z}_{m\mu}+\hat{\Omega}_m^{\ n}\widetilde{z}_{n\mu}.
\label{zmuB}
\end{eqnarray}

To maintain the vectors $l^\mu$ and $n^\mu$ real, the $\hat{\Sigma}_m^{\ n}$ matrix has to satisfy $\hat{\Sigma}_m^{\ n}\in\mathbb{R}$ and $\hat{\Sigma}_m^{\ 2}=\hat{\Sigma}_m^{*3}$ for $m,n=0,1$. Additionally, we require that $\hat{\Sigma}_2^{\ 3}=\hat{\Sigma}_3^{*2}$, $\hat{\Sigma}_2^{\ 2}=\hat{\Sigma}_3^{*3}$, and that $\hat{\Sigma}_2^{\ m}=\hat{\Sigma}_3^{*m}$ for $m=0,1$  in order to $m^{\mu}$ and $\bar{m}^\mu$ remain as complex conjugates of each other. To simplify notation we drop the hat off the perturbation terms $\hat{\Sigma}_m^{\ n},\hat{\Omega}_m^{\ n}$ and keep in mind through the rest of this section that now $\Sigma_m^{\ n}$ and $\Omega_m^{\ n}$ carry exclusively perturbed quantities. Then, the metric can be written as\footnote{In equation (\ref{guvab}) we have explicitly indicated there are second order terms of $\Omega_m^{\ n}$. From this point forward we will omit the second order dependency in every equation for compactness and, unless otherwise noted, every equal sign should be understood as such only to first order of $\Sigma$ or $\Omega$.}

\begin{equation}
g_{\mu\nu}=\gamma^{mn}z_{m\mu}z_{n\nu}=\widetilde{g}_{\mu\nu}+\Omega^{mn}(\widetilde{z}_{m\mu}\widetilde{z}_{n\nu}+\widetilde{z}_{m\nu}\widetilde{z}_{n\mu})+\mathcal{O}(\Omega^2),
\label{guvab}
\end{equation}

where $\Omega^{mn}=\gamma^{mp}\Omega_p^{\ n}$. Our first task will be to find a relation between $\Sigma_m^{\ n}$ and $\Omega_m^{\ n}$ such that the orthogonal properties of the tetrad formalism hold to first order of $\Sigma,\Omega$. Of course, these properties are assumed to be satisfied for the background tetrad. It is not difficult to prove that the relation we are looking for is $\Omega_m^{\ n}=-\gamma_{mp}\Sigma_q^{\ p}\gamma^{qn}$, or alternatively, $\Omega^{mn}=-\Sigma^{nm}$. Using this result one can next verify that $g^{\mu\rho}g_{\rho\nu}=\delta^\mu_{\ \nu}$, and so, the fundamental equations of the formalism are consistent.

With the tetrad given by equation (\ref{zmuB}) the quantities $\mathcal{Z}_{abc}$ related to the spin coefficients may be computed. However, note that the connection $\Gamma$ associated to the operator $\nabla$ appearing in these quantities is compatible with the metric $g$, not with the background metric $\widetilde{g}$. Naturally, the components of the connection $\Gamma$ can be expressed as $\Gamma^\rho_{\mu\nu}=\widetilde{\Gamma}^\rho_{\mu\nu}+\hat{\Gamma}^\rho_{\mu\nu}$. We obtain, thus,

\begin{equation}
\mathcal{Z}_{abc}=\widetilde{\mathcal{Z}}_{abc}-\hat{\Gamma}_{cab}+\widetilde{D}_a\Omega_{cb}+\widetilde{\mathcal{Z}}_{abp}\Omega_c^{\ p}+\widetilde{\mathcal{Z}}_{apc}\Sigma_b^{\ p}+\widetilde{\mathcal{Z}}_{pbc}\Sigma_a^{\ p},
\label{zabc}
\end{equation}

where we have defined $\hat{\Gamma}_{cab}=\widetilde{z}_{c\alpha}\hat{\Gamma}^\alpha_{\mu\nu}\widetilde{z}_a^{\ \mu}\widetilde{z}_b^{\ \nu}$. The components of the perturbed connection may be found by using the compatibility condition $\nabla_\alpha g_{\mu\nu}=0$ and the torsion free symmetry $\hat{\Gamma}_{abc}=\hat{\Gamma}_{acb}$. A straightforward, but somewhat long, calculation yields\footnote{Round brackets will be used to denote symmetrization of the indices enclosed, while square brackets to denote anti-symmetrization.}

\begin{equation}
\hat{\Gamma}_{abc}=\widetilde{D}_{(b}\Omega_{c)a}+\widetilde{D}_{[b}\Omega_{a]c}+\widetilde{D}_{[c}\Omega_{a]b}+\widetilde{\mathcal{Z}}_{(bc)p}\Xi_a^{\ p}+\widetilde{\mathcal{Z}}_{[ba]p}\Xi_c^{\ p}+\widetilde{\mathcal{Z}}_{[ca]p}\Xi_b^{\ p},
\label{christoffel}
\end{equation}

with $\Xi_m^{\ n}=\Omega_m^{\ n}-\Sigma_m^{\ n}$, and also $\Pi_m^{\ n}=\Omega_m^{\ n}+\Sigma_m^{\ n}$, which will be used in the next equation. Substituting (\ref{christoffel}) in (\ref{zabc}), and after some algebraic simplifications, we get $\mathcal{Z}_{abc}$ in terms only of background quantities and metric perturbations,

\begin{equation}
\mathcal{Z}_{abc}=\widetilde{\mathcal{Z}}_{abc}+\widetilde{D}_{[b}\Sigma_{a]c}-\widetilde{D}_{[c}\Sigma_{a]b}+\widetilde{D}_{[b}\Sigma_{c]a}+\Xi_{[b}^{\ m}\widetilde{\mathcal{Z}}_{c]am}+\widetilde{\mathcal{Z}}_{am[c}\Pi_{b]}^{\ m}+\Xi_a^{\ m}\widetilde{\mathcal{Z}}_{[cb]m}+\widetilde{\mathcal{Z}}_{mbc}\Sigma_a^{\ m}.
\label{zabc2}
\end{equation}

This equation is manifestly anti-symmetric in its last two indices as the quantity $\mathcal{Z}_{abc}$ should be. Though lengthy, equation (\ref{zabc2}) describes how the spin coefficients, which are necessary for the Newman-Penrose formalism, change to first order for any given perturbation $\Sigma_m^{\ n}$.

\subsection{Perturbed Tetrad Rotations}

Consider a transformation of the perturbation terms $\Omega^{mn}\rightarrow\Omega^{mn}+\Omega^{\prime mn}$. From (\ref{guvab}) it can be seen that

\begin{equation}
g_{\mu\nu}\rightarrow g_{\mu\nu}+\Omega^{\prime mn}(\widetilde{z}_{m\mu}\widetilde{z}_{n\nu}+\widetilde{z}_{m\nu}\widetilde{z}_{n\mu}).
\nonumber 
\end{equation}

Since the expression in parenthesis is symmetric in its tetrad indices, the metric will then be invariant under these type of transformations if we demand that $\Omega^{\prime mn}=-\Omega^{\prime nm}$. Not only the metric will be invariant, but naturally, also any other scalar or tensor derived from it, so long as the tensor does not possess tetrad indices. Therefore, there exists liberty in choosing the perturbation tetrad $\hat{z}_m^{\ \mu}=\Sigma_m^{\ n}\widetilde{z}_n^{\ \mu}$ since the $\Omega^{mn}$ that corresponds to a certain perturbed metric is not unique (recall that $\Omega_m^{\ n}=-\gamma_{mp}\Sigma_q^{\ p}\gamma^{qn}$). This of course is related to the group of Lorentz transformations that leave invariant the orthogonality properties of the formalism (see reference \cite{stephani}). However, for this case, the parameters of the Lorentz group should be taken as infinitesimal.

Under the transformation $\Omega^{mn}\rightarrow\Omega^{mn}+\Omega^{\prime mn}$, the previously defined $\Xi_m^{\ n}$ is invariant, while $$\Pi_m^{\ n}\rightarrow\Pi_m^{\ n}+2\Omega_m^{\prime n},$$ with $\Omega_m^{\prime n}=\gamma_{mp}\Omega^{\prime pn}$. Note that $\Omega^{\prime mn}=-\Omega^{\prime nm}$ implies that $\Omega^{\prime mn}=\Sigma^{\prime mn}$. Using these relations, we have that the quantities $\mathcal{Z}_{abc}$ transform as

\begin{equation}
\mathcal{Z}_{abc}\rightarrow\mathcal{Z}_{abc}+2\widetilde{\mathcal{Z}}_{am[c}\Omega_{b]}^{\prime m}+\widetilde{\mathcal{Z}}_{mbc}\Omega_a^{\prime m}.
\nonumber 
\end{equation}

Perhaps the most important benefit that the perturbed tetrad rotations provide lies in the differential operators $D_m$. They evidently change as $D_m\rightarrow D_m+\Sigma_m^{\prime n}\widetilde{D}_n$, but because there is some freedom in choosing the perturbation tetrad vectors, we may then conveniently pick them so that, for instance, $D_m=\widetilde{D}_m+\chi\widetilde{D}_n$ for some fixed $m\neq n$, and a scalar field $\chi$. In the following section we take advantage of this particular property, simplifying thus our calculations.

It is important to notice that, when performing any rotation through $\Omega^{mn}\rightarrow\Omega^{mn}+\Omega^{\prime mn}$, one has to be careful that the rotated vectors $l^\mu$ and $n^\mu$ end up being real, and that $m^{\mu}$ and $\bar{m}^\mu$ remain as complex conjugates. This restricts the possible valid rotations that can be done. Taking into account these restrictions, one can be convinced that there is a total of six degrees of freedom, which is consistent with the fact that the group of Lorentz transformations is a six parameter group.

\section{III. Gravitational Perturbations in Spherically Symmetric Space-Times}

For the remainder of this work we focus on four-dimensional stationary and spherically symmetric space-times $(M, g_{\mu\nu})$ whose line element, without loss of generality, can be written in the form

\begin{equation}
ds^2=g_0(r)dt^2-g_1(r)dr^2-g_2(r)d\Omega^2,
\label{ds2}
\end{equation}

where we have introduced a radial coordinate $r$ and the metric elements $g_{0,1,2}(r)$, which are arbitrary functions of said coordinate. Also, $d\Omega^2$ is the standard metric on the two-sphere. An orthonormal frame for this metric is simply given by

\begin{equation}
X_0=\frac{1}{\sqrt{g_0}}\partial_t, \hspace{4mm} X_1=\frac{1}{\sqrt{g_1}}\partial_r, \hspace{4mm} X_2=\frac{1}{\sqrt{g_2}}\partial_\theta, \hspace{4mm} X_3=\frac{1}{\sqrt{g_2}\sin\theta}\partial_\varphi.
\label{Xmu}
\end{equation}

From frame (\ref{Xmu}), a null tetrad can be constructed by taking appropriate linear combinations of the $X$ vectors. In this paper we will take advantage of the symmetries of the space-time, namely the fact that $\partial_t$ and $\partial_\varphi$ are Killing vectors, and choose $\widetilde{l}^\mu$ and $\widetilde{n}^\mu$ so that they lie in the subspace spanned by said Killing vectors. This can also be extended to axially symmetric space-times. Hence, the vectors of the null tetrad will be\footnote{This choice of tetrad differs from the usual, for instance in \cite{hamiltonNP}, in which $\widetilde{l}^\mu$ and $\widetilde{n}^\mu$ are combination of the $X_0$ and $X_1$ vectors. Our choice will come with certain advantages that will later be seen.}

\begin{equation}
\widetilde{l}^\mu=\frac{1}{\sqrt{2}}(X_0^\mu+X_3^\mu), \hspace{4mm} \widetilde{n}^\mu=\frac{1}{\sqrt{2}}(X_0^\mu-X_3^\mu), \hspace{4mm} \widetilde{m}^\mu=\frac{1}{\sqrt{2}}(X_1^\mu+iX_2^\mu).
\label{tetrada}
\end{equation} 

A direct evaluation of the spin coefficients of the Newman-Penrose formalism with metric (\ref{ds2}) and tetrad (\ref{tetrada}) yields that the only non-vanishing coefficients are $\widetilde{\kappa}$, $\widetilde{\nu}$, $\widetilde{\tau}$, $\widetilde{\pi}$, $\widetilde{\alpha}$ and $\widetilde{\beta}$. Additionally, the following properties hold 

\begin{equation}
\widetilde{\kappa}+\widetilde{\nu}^*=\widetilde{\tau}+\widetilde{\pi}^*=\widetilde{\alpha}+\widetilde{\beta}=0, \hspace{4mm} \widetilde{\kappa}+\widetilde{\nu}=-\widetilde{\tau}-\widetilde{\pi}, \hspace{4mm} \widetilde{\alpha}=\frac{1}{4}\left(\widetilde{\nu}+\widetilde{\nu}^*+\widetilde{\tau}+\widetilde{\tau}^*\right),
\label{id}
\end{equation}

with $\widetilde{\alpha},\widetilde{\beta}\in\mathbb{R}$. Notice that as a consequence of our choice of vectors $\widetilde{l}^\mu$ and $\widetilde{n}^\mu$ we will have that $\widetilde{D}\widetilde{\phi}=\widetilde{\Delta}\widetilde{\phi}=0$ for any background scalar quantity $\widetilde{\phi}$, including these spin coefficients.

We now add a perturbation term $h_{\mu\nu}$ to the background metric introduced in this section. Following the pioneering work of Regge and Wheeler \cite{rw} we consider a perturbation of the form 

\begin{eqnarray}
h_{\mu\nu}=
\begin{bmatrix}
	0 & 0 & 0 & h_0 \\
	0 & 0 & 0 & h_1 \\
	0 & 0 & 0 & 0 \\
	h_0 & h_1 & 0 & 0 \\ 	
\end{bmatrix},
\label{huv}
\end{eqnarray}

with $h_{\mu\nu}$ expressed in the coordinate basis $\left\{t,r,\theta,\varphi\right\}$. Regge and Wheeler obtained this particular (and simple) expression for $h_{\mu\nu}$ through a gauge transformation of the most general perturbation whose angular part consists of products of scalar, vector, and tensor spherical harmonics $Y_{\ell,m}(\theta,\varphi)$ on the 2-sphere. This gauge is sometimes called the Regge-Wheeler gauge. Furthermore, (\ref{huv}) represents a perturbation of $(-1)^{\ell+1}$ parity, which Regge and Wheeler named as odd, due to its negative symmetry under reflections about the origin. This type of perturbation is sometimes also referred to as axial. The $\varphi$-dependence of the perturbation can be eliminated without significant loss of information by setting $m=0$. This is possible since the background space-time is spherically symmetric. Therefore, we have that $h_{0,1}=h_{0,1}(t,r,\theta)$.

Using the one-forms of the background tetrad $\left\{\widetilde{l}_\mu,\widetilde{n}_\mu,\widetilde{m}_\mu,\widetilde{\bar{m}}_\mu\right\}$ as a basis, we can write 

\begin{equation}
h_{\mu\nu}=f_0(\widetilde{n}_\mu\widetilde{n}_\nu-\widetilde{l}_\mu\widetilde{l}_\nu)+2f_1[\widetilde{l}_{(\mu}\widetilde{m}_{\nu)}+\widetilde{l}_{(\mu}\widetilde{\bar{m}}_{\nu)}-\widetilde{n}_{(\mu}\widetilde{m}_{\nu)}-\widetilde{n}_{(\mu}\widetilde{\bar{m}}_{\nu)}],
\nonumber 
\end{equation}

where $f_0=h_0/\sqrt{g_0g_2}\sin\theta$ and $f_1=h_1/2\sqrt{g_1g_2}\sin\theta$. It can be verified that an acceptable tetrad for the perturbed metric $g_{\mu\nu}=\widetilde{g}_{\mu\nu}+h_{\mu\nu}$ is given by

\begin{equation}
l_{\mu}=\widetilde{l}_{\mu}+\frac{1}{2}f_0\widetilde{n}_\mu-f_1(\widetilde{m}_\mu+\widetilde{\bar{m}}_\mu), \hspace{4mm} n_{\mu}=\widetilde{n}_{\mu}-\frac{1}{2}f_0\widetilde{l}_\mu+f_1(\widetilde{m}_\mu+\widetilde{\bar{m}}_\mu), \hspace{4mm} m_{\mu}=\widetilde{m}_{\mu},
\label{tetradaper}
\end{equation}

from which the elements of $\Omega_m^{\ n}$ can be easily read off

\begin{eqnarray}
\Omega_m^{\ n}=
\begin{bmatrix}
	0 & f_0/2 & -f_1 & -f_1 \\
	-f_0/2 & 0 & f_1 & f_1 \\
	0 & 0 & 0 & 0 \\
	0 & 0 & 0 & 0 \\ 	
\end{bmatrix}.
\nonumber 
\end{eqnarray}

It will more helpful, though, to represent the perturbation in terms of $\Sigma_m^{\ n}=-\gamma_{mp}\Omega_q^{\ p}\gamma^{qn}$, obtaining thus 

\begin{eqnarray}
\Sigma_m^{\ n}=
\begin{bmatrix}
	0 & -f_0/2 & 0 & 0 \\
	f_0/2 & 0 & 0 & 0 \\
	f_1 & -f_1 & 0 & 0 \\
	f_1 & -f_1 & 0 & 0 \\ 	
\end{bmatrix}.
\label{sigma}
\end{eqnarray}

As it was already stated in the previous section, any given metric perturbation does not uniquely define $\Omega_m^{\ n}$, and consequently, $\Sigma_m^{\ n}$. We will be interested in a perturbed tetrad such that 

\begin{equation}
D\widetilde{\phi}=(\chi_1\widetilde{D}+\chi_2\widetilde{\Delta})\widetilde{\phi}=0, \hspace{4mm} \Delta\widetilde{\phi}=(\xi_1\widetilde{D}+\xi_2\widetilde{\Delta})\widetilde{\phi}=0,
\label{D1D2}
\end{equation}

where again, $\widetilde{\phi}$ is any background scalar quantity and $\chi_{1,2}$, $\xi_{1,2}$ are elements of $\Sigma_m^{\ n}$. It turns out that precisely the matrix given by (\ref{sigma}) describes the perturbation tetrad with this desired property. Nonetheless, it is important to mention that one can always find, through an adequate tetrad rotation, a perturbation tetrad such that (\ref{D1D2}) holds in a spherically symmetric (even in an axially symmetric, for that matter) background space-time. This is possible too due to our previous election of background vectors $\widetilde{l}^\mu$ and $\widetilde{n}^\mu$, namely, the fact that they lie in the subspace spanned by Killing vectors. Another advantage that this $\Sigma_m^{\ n}$ possess is that $\delta\widetilde{\phi}=\widetilde{\delta}\widetilde{\phi}$. However, this will not always be the case for an arbitrary metric perturbation, even performing a perturbed tetrad rotation. Additionally it can be seen that $\widetilde{D}f_{0,1}=\widetilde{\Delta}f_{0,1}$ because of the $\varphi$ independence of those functions.

With finally an explicit expression for the perturbation matrix $\Sigma_m^{\ n}$, we can proceed to compute the perturbed spin coefficients using equation (\ref{zabc2}). Our objective then will be to write the components of the Ricci tensor $R_{\mu\nu}$ in terms of these spin coefficients using the equations of the Newman-Penrose formalism. It can already be foreseen that we will obtain second-order partial differential equations for the perturbation functions $f_0$ and $f_1$ due to the fact that the formalism provides first-order partial differential equations for the spin coefficients, and these in turn, have first-order derivatives of said functions. Since the calculation of the NP quantities is pretty much straightforward and the results are numerous, they will be shown separately in Appendix A, and we should cite them in the following as needed.

In the Newman-Penrose formalism, 10 curvature related quantities $\Phi_{AB}$ ($A,B=0,1,2$), and $\Lambda=R/24$, are defined. These are merely the projection of the tetrad vectors into the Ricci tensor, i.e., $R_{\mu\nu}z_m^{\ \mu}z_n^{\ \nu}$, and a rescaling of the Ricci scalar $R$, respectively. See equations (NP 4.3b) for their explicit expressions. By either a direct calculation of these quantities, or by the vanishing of the background spin coefficients, the following holds for the background metric

\begin{equation}
\widetilde{\Phi}_{01}=\widetilde{\Phi}_{10}=\widetilde{\Phi}_{12}=\widetilde{\Phi}_{21}=0, \hspace{4mm} \widetilde{\Phi}_{00}=\widetilde{\Phi}_{22}.
\label{PhiA}   
\end{equation}  

This also follows from the fact that the background Ricci tensor admits the form $\widetilde{R}_{\mu\nu}=diag\left[\widetilde{R}_{00},\widetilde{R}_{11},\widetilde{R}_{22},\widetilde{R}_{33}\right]$ in the $\{t,r,\theta,\varphi\}$ basis. Using (NP 4.2a), (NP 4.2n) and (\ref{id}), it can be seen that the last equality in (\ref{PhiA}) implies that 

\begin{equation}
\widetilde{\delta}\widetilde{\nu}+\widetilde{\delta^*}\widetilde{\kappa}=2\widetilde{\alpha}(\widetilde{\nu}+\widetilde{\kappa}).
\label{id2}
\end{equation}

It is important to realize that, apart from $\Lambda$, the $\Phi_{AB}$ quantities depend manifestly on the tetrad choice. Even upon fixing the background tetrad, $\Phi_{AB}$ will vary with perturbed rotations such as the ones described in the previous section. Since the Ricci tensor itself is invariant to these type of transformations, we look for expressions of its components in the coordinate basis and constructed from the quantities $\Phi_{AB}$ and $\Lambda$. More precisely, we will look for the components of $R_{\mu\nu}$ in the orthonormal frame (\ref{Xmu}). 

In terms of the background tetrad basis, the orthonormal basis can be written as $X_\alpha^{\ \mu}=\widetilde{\Gamma}_\alpha^{\ m}\widetilde{z}_n^{\ \mu}$. From (\ref{tetrada}) it can be easily seen that, $$\widetilde{\Gamma}_\alpha^{\ m}=\frac{1}{\sqrt{2}}\begin{bmatrix}
	1 & 1 & 0 & 0 \\
	0 & 0 & 1 & 1 \\
	0 & 0 & -i & i \\
	1 & -1 & 0 & 0 \\ 	
\end{bmatrix}.$$

Similarly, in terms of the perturbed tetrad we have that $X_\alpha^{\ \mu}=\Gamma_\alpha^{\ m}z_m^{\ \mu}$. Using the fact that $z_m^{\ \mu}=(\delta_m^{\ n}+\Sigma_m^{\ n})\widetilde{z}_n^{\ \mu}$, we obtain $\Gamma_\alpha^{\ n}=\widetilde{\Gamma}_\alpha^{\ m}(\delta_m^{\ n}-\Sigma_m^{\ n})$ to first order in $\Sigma$, or explicitly,

\begin{eqnarray}
\Gamma_\alpha^{\ n}=\frac{1}{\sqrt{2}}
\begin{bmatrix}
	1-f_0/2 & 1+f_0/2 & 0 & 0 \\
	-2f_1 & 2f_1 & 1 & 1 \\
	0 & 0 & -i & i \\
	1+f_0/2 & -1+f_0/2 & 0 & 0 \\ 	
\end{bmatrix}.
\label{Gamma}
\end{eqnarray}

We can then write,

\begin{align}
X_0^\mu&=\frac{1}{\sqrt{2}}\left[\left(1-\frac{f_0}{2}\right)l^\mu+\left(1+\frac{f_0}{2}\right)n^\mu\right], & X_1^\mu&=\frac{1}{\sqrt{2}}\left[m^\mu+\bar{m}^\mu+2f_1(n^\mu-l^\mu)\right], \nonumber\\
X_2^\mu&=\frac{i}{\sqrt{2}}\left[\bar{m}^\mu-m^\mu\right], & X_3^\mu&=\frac{1}{\sqrt{2}}\left[\left(1+\frac{f_0}{2}\right)l^\mu-\left(1-\frac{f_0}{2}\right)n^\mu\right]. 
\label{Xmu2}
\end{align}

The $X$ vectors of equation (\ref{Xmu2}) can be shown to be invariant under transformations $\Sigma_m^{\ n}\rightarrow\Sigma_m^{\ n}+\Sigma_m^{\prime n}$, and to first order in $\Sigma$, by noting that $\Gamma_\alpha^{\ n}\rightarrow\Gamma_\alpha^{\ n}-\widetilde{\Gamma}_\alpha^{\ m}\Sigma_m^{\prime n}$, and $z_m^{\ \mu}\rightarrow z_m^{\ \mu}+\Sigma_m^{\prime n}\widetilde{z}_n^{\ \mu}$. Thus, we may find the desired invariant equations for the Ricci components by contracting these vectors with the Ricci tensor field. Unfortunately, this has to be done for the 10 independent components of said tensor, yielding the following relations

\begin{align}
\hat{\mathcal{R}}_{00}&=-\hat{\Phi}_{00}-\hat{\Phi}_{22}+2(3\hat{\Lambda}-\hat{\Phi}_{11}), & \hat{\mathcal{R}}_{01}&=-\hat{\Phi}_{01}-\hat{\Phi}_{10}-\hat{\Phi}_{12}-\hat{\Phi}_{21}, \nonumber\\
\hat{\mathcal{R}}_{02}&=i(\hat{\Phi}_{01}-\hat{\Phi}_{10}+\hat{\Phi}_{12}-\hat{\Phi}_{21}), & \hat{\mathcal{R}}_{03}&=-\hat{\Phi}_{00}+\hat{\Phi}_{22}+2(3\widetilde{\Lambda}-\widetilde{\Phi}_{11})f_0, \nonumber\\
\hat{\mathcal{R}}_{11}&=-\hat{\Phi}_{02}-\hat{\Phi}_{20}-2(3\hat{\Lambda}+\hat{\Phi}_{11}), & \hat{\mathcal{R}}_{12}&=i(\hat{\Phi}_{02}-\hat{\Phi}_{20}), \nonumber\\
\hat{\mathcal{R}}_{13}&=-\hat{\Phi}_{01}+\hat{\Phi}_{12}-\hat{\Phi}_{10}+\hat{\Phi}_{21}+4(3\widetilde{\Lambda}-\widetilde{\Phi}_{11}+\widetilde{\Phi}_{00})f_1, & \hat{\mathcal{R}}_{22}&=\hat{\Phi}_{02}+\hat{\Phi}_{20}-2(3\hat{\Lambda}+\hat{\Phi}_{11}), \nonumber\\
\hat{\mathcal{R}}_{23}&=i(\hat{\Phi}_{01}-\hat{\Phi}_{10}+\hat{\Phi}_{21}-\hat{\Phi}_{12}), & \hat{\mathcal{R}}_{33}&=-\hat{\Phi}_{00}-\hat{\Phi}_{22}-2(3\hat{\Lambda}-\hat{\Phi}_{11}),
\label{Ricci}
\end{align}

where we have defined $\hat{\mathcal{R}}_{\alpha\beta}=\hat{R}_{\mu\nu}X_\alpha^{\ \mu}X_\beta^{\ \nu}$. In equations (\ref{Ricci}) we have written only the perturbation terms (denoted by a hat), that is, the terms of first order in $f_{0,1}$. Naturally, the background terms that should appear on both sides of the equations, which are of order zero, cancel each other out. Hereafter, we drop the tilde off the background quantities and so, any quantity or operator without a hat should be understood to be of the background space-time, except for the perturbation functions $f_0$ and $f_1$ (same convention as Appendix A).

Taking the results (\ref{PhiB}) from Appendix A, one can realize that the only non-vanishing components of $\hat{\mathcal{R}}_{\alpha\beta}$ are

\begin{eqnarray}
\hat{\mathcal{R}}_{03}&=&2(\delta_+-6\alpha)Df_1+\left[(\delta_-+2\kappa_+)\delta_--(\delta_++\kappa_-+3\pi_-)\delta_++4(\kappa_-^2-\kappa_+^2)+2(3\Lambda-\Phi_{11})\right]f_0, \nonumber\\
\hat{\mathcal{R}}_{13}&=&2\left[D^2+(\delta_-+4\kappa_+)(\delta_--2\kappa_+)+2(3\Lambda-\Phi_{11}+\Phi_{00})\right]f_1-(\delta_+-3\kappa_-+\pi_-)Df_0, \nonumber\\
\hat{\mathcal{R}}_{23}&=&i(\delta_--2\kappa_+)Df_0-2i(\delta_++\kappa_-+3\pi_-)(\delta_--2\kappa_+)f_1.
\label{R3}
\end{eqnarray}

With the help of the commutator $[\delta_--2\kappa_+,\delta_+]=(\kappa_-+\pi_-)(\delta_-+2\kappa_+)$, the $\hat{\mathcal{R}}_{23}$ component of the past equations can be rewritten as $$\hat{\mathcal{R}}_{23}=i(\delta_--2\kappa_+)\left[Df_0-2(\delta_++2\pi_-)f_1\right].$$

So far we have focused on describing how the space-time geometry is modified when adding a small term to the background metric. In fact, equations (\ref{R3}) describe precisely this change up to linear order. To obtain a set of suitable perturbation equations, however, one must take into account both sides of the field equations that a particular gravitational source yields. Namely, equations (\ref{R3}) will give information of one side of the field equations, whereas the other side will be determined by the physical variables of the source. This gravitational source may in principle be also perturbed.

As an initial approach, consider background field equations that consist of a simple structure such as

\begin{equation}
R_{\mu\nu}=S_{\mu\nu}+Sg_{\mu\nu},
\label{Rmat}
\end{equation}

where $S_{\mu\nu}$ is a symmetrical tensor and $S$ a scalar, both containing the physical properties of the source. Spherical symmetry, combined with expression (\ref{Rmat}) for the Ricci tensor, imposes that the tensor $S_{\mu\nu}$ must be written as

\begin{equation}
S_{\mu\nu}=diag[S_{00}(r),S_{11}(r),S_{22}(r),S_{22}(r)\sin^2\theta]
\label{Suv}
\end{equation}

in the $\{t,r,\theta,\varphi\}$ basis. In this paper we will treat the simple case in which $S_{\mu\nu}$ and $S$ need not be perturbed in order to solve the modified field equations. Hence, 

\begin{equation}
\hat{R}_{\mu\nu}=Sh_{\mu\nu}.
\label{RS}
\end{equation}

We should remark here that by leaving unchanged the physical parameters that appear in $S_{\mu\nu}$ and $S$, perturbations in space-times that solve the Einstein-Maxwell equations are, in the most common case, excluded from the treatment that is described in the following. The reason for this is that the set of Maxwell's equations fall beyond the scope of this paper. Hence, they are not guaranteed to hold up to liner order when perturbing the metric without adding a small term in the electromagnetic tensor $F_{\mu\nu}$ too. The same applies to electromagnetic perturbations. Examples of valid space-times and field equations will be provided later.

When contracting the necessary $X$ vectors with the perturbed Ricci tensor (\ref{RS}), the non-vanishing components of the field equations become

\begin{eqnarray}
2(\delta_+-6\alpha)Df_1+\left[(\delta_-+2\kappa_+)\delta_--(\delta_++\kappa_-+3\pi_-)\delta_++4(\kappa_-^2-\kappa_+^2)+2(3\Lambda-\Phi_{11})-S\right]f_0&=&0, \nonumber\\
2\left[D^2+(\delta_-+4\kappa_+)(\delta_--2\kappa_+)+2\Lambda_s-S\right]f_1-(\delta_+-3\kappa_-+\pi_-)Df_0&=&0, \nonumber\\
i(\delta_--2\kappa_+)\left[Df_0-2(\delta_++2\pi_-)f_1\right]&=&0,
\label{R3mat}
\end{eqnarray}

with $\Lambda_s=3\Lambda-\Phi_{11}+\Phi_{00}$. We may now attempt to solve the system of equations (\ref{R3mat}). Consider the last equation, we have already factored it in a way that can be easily solved. Notice that the expression in parentheses cannot vanish since $\delta_-$ is a differential operator and $\kappa_+$ is a scalar quantity. Thus, the left side of this equation will vanish if the expression in square brackets also does, or by the application of the operator in parentheses to the quantity in square brackets. We will examine the first possibility, that is,

\begin{equation}
Df_0=2(\delta_++2\pi_-)f_1.
\label{Df0}
\end{equation}

By inserting (\ref{Df0}) in the $\hat{\mathcal{R}}_{13}$ component of (\ref{R3mat}), an equation for the perturbation function $f_1$ can finally be found,

\begin{equation}
\left[D^2+(\delta_-+4\kappa_+)(\delta_--2\kappa_+)-(\delta_+-3\kappa_-+\pi_-)(\delta_++2\pi_-)+2\Lambda_s-S\right]f_1=0.
\label{f1}
\end{equation}

We are left, though, with the first equation in (\ref{R3mat}) yet to be solved with the inconvenient that the perturbation functions $f_0$ and $f_1$ have already been used to satisfy the other two equations in the system. The $\hat{\mathcal{R}}_{03}$ component can be shown to vanish only if 

\begin{equation}
(\delta_+-4\alpha)(S-2\Lambda_s)=0.
\label{R03c}
\end{equation}

This in turn implies that $S_{22}(r)=c$ in equation (\ref{Suv}), being $c$ an integration constant (see Appendix B for details). Unfortunately at this point, we are forced to abandon the generality that has been conserved until now regarding the spherically symmetric space-times here considered, and restrict ourselves to those in which the $S_{22}$ component is a constant. Examples that fulfill this condition, besides vacuum space-times, are the solutions of the Einstein-scalar field equations with a self-interacting potential $R_{\mu\nu}=\pm\nabla_\mu\phi\nabla_\nu\phi+\mathcal{V}(\phi)g_{\mu\nu}$, as well as perfect fluid solutions.  For both of these cases, $c=0$. Many wormhole space-times arise as solutions to the first type of field equations, hence, our results can be applied to them.

For reasons explained in the next section, we shall opt to replace the perturbation function $f_1$ with $Q=2\sqrt{g_0}\sin\theta f_1$. To do so, the following helpful relations can be verified to be true by examining the spin coefficients and operators of (\ref{spinoppm}),

\begin{equation}
\delta_-\left(\frac{1}{\sin\theta}\right)=-\frac{2\kappa_+}{\sin\theta}, \hspace{4mm} \delta_+\left(\frac{1}{\sqrt{g_0}}\right)=\frac{\kappa_--\pi_-}{\sqrt{g_0}}.
\nonumber
\end{equation}

Substituting $f_1=Q/2\sqrt{g_0}\sin\theta$ in (\ref{f1}), and employing these two equalities, we at last arrive to our master equation for odd-parity perturbations,

\begin{equation}
\left[D^2+(\delta_-+2\kappa_+)(\delta_--4\kappa_+)-(\delta_+-2\kappa_-)(\delta_++\pi_-+\kappa_-)+2\Lambda_s-S\right]Q=0.
\label{Q}
\end{equation} 

The notation introduced throughout the paper allows us to easily identify the terms appearing in the master equation. The $D$ operator is associated to the time dependence of the perturbation, the second term is associated with the angular part due to it containing the $\delta_-$ operators, and the third term is related to the radial part because of the $\delta_+$ operators. In (\ref{Q}) there also appears a background matter term which is purely radial. It is natural then to propose a separable ansatz of the form $Q=T(t)R(r)\Theta(\theta)$. With such a proposed solution, the angular part of the master equation will yield the following differential equation when inserting the explicit expressions for the spin coefficients and operators,

\begin{equation}
\frac{d^2\Theta}{d\theta^2}-\frac{1}{\tan\theta}\frac{d\Theta}{d\theta}=-\ell(\ell+1)\Theta.
\label{ptheta}
\end{equation}

Equation (\ref{ptheta}) has for solution $\Theta_\ell=\sin\theta dP_\ell(\cos\theta)/d\theta$, where $P_\ell(\cos\theta)$ are the well-known Legendre polynomials. This result was of course, expected, owing to the spherical symmetry of the line element (\ref{ds2}) and to the decomposition in tensor spherical harmonics of the perturbation that Regge and Wheeler previously used. In fact, this part of the solution is, obviously, the same that appears in their paper. 

The radial part of the master equation is obtained by computing all of the terms appearing in (\ref{Q}), this is,

\begin{eqnarray}
\frac{1}{g_0}\frac{\partial^2Q}{\partial t^2}-\frac{1}{g_2}\left(\frac{\partial^2Q}{\partial\theta^2}-\frac{1}{\tan\theta}\frac{\partial Q}{\partial\theta}+2Q\right)&-&\frac{1}{\sqrt{g_1}}\frac{\partial}{\partial r}\left(\frac{1}{\sqrt{g_1}}\frac{\partial Q}{\partial r}\right)-\frac{g'_0}{2g_0g_1}\frac{\partial Q}{\partial r} \nonumber\\
&+&\sqrt{\frac{g_2}{g_0g_1}}\frac{d}{dr}\left[\sqrt{\frac{g_0}{g_1}}\frac{d}{dr}\left(\frac{1}{\sqrt{g_2}}\right)\right]Q+2(2\Lambda_s-S)Q=0.
\nonumber
\end{eqnarray}

The above second-order partial differential equation can be further simplified by considering the previously introduced ansatz for $Q$, whose angular part has already been solved, and with the additional assumption of an harmonic dependence on time, i.e., $Q=e^{i\omega t}R(r)\sin\theta dP_l(\cos\theta)/d\theta$. Furthermore, introducing the tortoise coordinate $r_*$ defined by

\begin{equation}
\frac{d}{dr_*}=\sqrt{\frac{g_0}{g_1}}\frac{d}{dr},
\nonumber
\end{equation}

and substituting equations (\ref{const}, \ref{ptheta}), the master equation can finally be rewritten in a very compact form as,

\begin{equation}
\frac{d^2R}{dr_*^2}-\left(V(r)-\omega^2\right)R=0,
\label{Qrad}
\end{equation}

with 

\begin{equation}
V(r)=\frac{g_0}{g_2}[\ell(\ell+1)-2(c+1)]+\sqrt{g_2}\frac{d^2}{dr_*^2}\left(\frac{1}{\sqrt{g_2}}\right).
\label{V}
\end{equation}

We have reduced the master equation to an eigenvalue problem for $\omega^2$ and the operator $\mathcal{H}=-d^2/dr_*^2+V(r)$, which is linear and self-adjoint. An operator of this type is sometimes called a Schr\"odinger operator with effective potential $V(r)$. The potential found here can now be compared to some previous results. For instance, it can be verified that this potential reduces to that of Regge-Wheeler when inserting the corresponding metric components of the Schwarzschild metric. It also coincides with the potential of axial and uncharged perturbations for electrically neutral background space-times found in \cite{phantom1}. The potential of our paper can be seen as a slight generalization of the uncharged result to arbitrary spherically symmetric line elements\footnote{In \cite{phantom1} the space-times considered are supported by a phantom scalar field (hence, $c=0$ in equation (\ref{V})), with the assumption that $g_{tt}=-1/g_{rr}$ for the metric tensor.} that admit field equations of the type (\ref{Rmat}), with the exception of Einstein-Maxwell solutions. During the next sections of this work we will analyze the potential $V(r)$ of some wormhole examples, along with their properties.

However, before ending this section it might be worth discussing the perturbed stress-energy tensor $T_{\mu\nu}+\hat{T}_{\mu\nu}$ and its conservation law. From (\ref{Rmat}), its background and perturbation terms are respectively given in geometrized units ($G=c=1$) by 

\begin{equation}
8\pi T_{\mu\nu}=S_{\mu\nu}-\left(\frac{1}{2}s+S\right)g_{\mu\nu}, \hspace{5mm} 8\pi\hat{T}_{\mu\nu}=-\left(\frac{1}{2}s+S\right)h_{\mu\nu},
\nonumber
\end{equation}

where $s=S_{\mu\nu}g^{\mu\nu}$. Note that $\hat{s}=S_{\mu\nu}h^{\mu\nu}=0$. It can be verified then that $\nabla_\mu(T^{\mu\nu}+\hat{T}^{\mu\nu})=0$ up to linear order of the perturbation, where the covariant differentiation $\nabla$ is compatible with the perturbed metric. This was expected, of course, as a result of solving in a consistent way the system (\ref{R3mat}) of field equations. For the case of the Einstein-scalar field equations with a self-interacting potential $\mathcal{V}(\phi)$, we have that $S_{\mu\nu}=\pm\nabla_\mu\phi\nabla_\nu\phi$ and $S=\mathcal{V}(\phi)$. It can be shown that the Klein-Gordon equation is implied by the conservation law of an arbitrary scalar field $\phi$ and hence, $$\pm\nabla^\mu\nabla_\mu\phi=\frac{d\mathcal{V}(\phi)}{d\phi},$$ up to first order of the perturbation too, and as long as the scalar field is not perturbed since then $\hat{S}_{\mu\nu}=\hat{S}=0$. This is the assumption made to solve the field equations.

\section{IV. Physical Regularity of the Perturbation}

In order for the gravitational perturbation to be of any physical relevance, it has to display an "acceptable" behavior throughout space-time, or at least asymptotically. One might naturally impose the condition that the metric perturbation functions of $h_{\mu\nu}$ do not grow without bound as $r\rightarrow\infty$ and deem that as physical regularity. Nevertheless, due to the gauge freedom that exists in General Relativity, this condition is not quite precise. Fortunately, the Newman-Penrose formalism can also be used to describe more accurately what this acceptable behavior is expected to be by means of the so-called ``peeling theorem'' \cite{np}. 

Consider the following vectors tangent to ingoing and outgoing radial null geodesics of the background metric,

\begin{equation}
k_\pm^\mu=X_0^\mu\pm X_1^\mu=\frac{1}{\sqrt{2}}\left(\widetilde{l}^\mu+\widetilde{n}^\mu\pm\widetilde{m}^\mu\pm\widetilde{\bar{m}}^\mu\right).
\nonumber 
\end{equation}

The next null rotations of our initial tetrad yield a new one such that the unperturbed part of $l''^\mu$ and $n''^\mu$ is aligned to the $k_+^\mu$ and $k_-^\mu$ vectors, respectively,

\begin{align}
l'^\mu=&l^\mu+a_1\bar{m}^\mu+a_1^*m^\mu+\left\|a_1\right\|^2n^\mu, \hspace{4mm} m'^\mu=m^\mu+a_1n^\mu, \hspace{4mm} n'^\mu=n^\mu, \nonumber\\
n''^\mu=&n'^\mu+a_2\bar{m}'^\mu+a_2^*m'^\mu+\left\|a_2\right\|^2l'^\mu,  \hspace{4mm} l''^\mu=l'^\mu, 
\label{rot12} 
\end{align}

with $a_1=1$ and $a_2=-1/2$. In equations (\ref{rot12}) and (\ref{weyl}), and only in those equations, we temporarily restore the convention of section II in which any given quantity $\xi$ of the space-time is written as the sum of a background term and a perturbation term, i.e., $\xi=\widetilde{\xi}+\hat{\xi}$. Under rotations (\ref{rot12}), the transformation laws of the Weyl scalars we will need are (see reference \cite{stephani})

\begin{align}
\psi'_0=\psi_0+4a_1\psi_1+6a_1^2\psi_2+4a_1^3\psi_3+a_1^4\psi_4, \hspace{4mm} \psi'_1=&\psi_1+3a_1\psi_2+3a_1^2\psi_3+a_1^3\psi_4, \hspace{4mm} \psi'_2=\psi_2+2a_1\psi_3+a_1^2\psi_4, \nonumber\\
\psi''_2=&\psi'_2+2a_2^*\psi'_1+a_2^{*2}\psi'_0.
\label{weyl}
\end{align}

If $a_1=1$ and $a_2=-1/2$, then $\psi''_2=\psi_0/4-\psi_2/2+\psi_4/4$. After substituting the expressions found in (\ref{psi}), the perturbed part of this Weyl scalar reduces to

\begin{equation}
\hat{\psi}''_2=\frac{1}{2}(\delta_-+2\kappa_+)\left[(\delta_+-2\kappa_-)f_0-2Df_1\right].
\label{psi2}
\end{equation} 

The physical significance of $\hat{\psi}''_2$ can be revealed by applying the operator $D$ to (\ref{psi2}), and then reducing it accordingly with some of the relations of the formalism here derived along with the master equation, thus obtaining

\begin{equation}
D\hat{\psi}''_2=\frac{1}{2\sqrt{g_0}\sin\theta}\delta_-\left[(\delta_-+2\kappa_+)(\delta_--4\kappa_+)+2\Lambda_s-S)\right]Q.
\label{Dpsi2}
\end{equation}

We have already solved the angular part of the master equation whose terms appear again in (\ref{Dpsi2}). By making use of said solution and some properties of the Legendre equation, the past expression can be rewritten as

\begin{equation}
\frac{\partial\hat{\psi}''_2}{\partial t}=-\frac{i\ell(\ell+1)}{g_2^{3/2}}\left[(\ell-1)(\ell+2)-c\right]T(t)R(r)P_\ell(\cos\theta),
\label{dtpsi2c}
\end{equation}

where we have made use of restriction (\ref{const}) too. In the case of space-times that solve the Einstein-scalar field equations (and vacuum space-times too) we have that $c=0$, and the meaning of 

\begin{equation}
\frac{\partial\hat{\psi}''_2}{\partial t}=-\frac{i(\ell+2)!}{g_2^{3/2}(\ell-2)!}T(t)R(r)P_\ell(\cos\theta)
\label{dtpsi2}
\end{equation}

becomes clearer, as well as the reason behind the use of the perturbation function $Q$. The peeling theorem establishes that the Weyl scalar $\psi_2$ asymptotically decays at null infinity as $1/\lambda'^3$, where $\lambda'$ is the affine parameter of a null geodesic that reaches said infinity. Considering the background radial null geodesics to which the unperturbed part of $l''^\mu$ and $n''^\mu$ are tangent to, we have that asymptotically $\lambda'\sim r$, due to $r$ being an appropriate radial coordinate. Asymptotically too, the metric component appearing in (\ref{dtpsi2}) goes as $g_2(r)\sim r^2$. The perturbation function $Q=T(t)R(r)\Theta_\ell(\theta)$, hence, manifestly describes the peeling property that the $\psi_2$ scalar should display at null infinity. From this analysis we can state that a regular behavior of $Q$ is one that does not alter the $1/r^3$ decay of the Weyl scalar $\hat{\psi}''_2$ when $r\rightarrow\infty$. Also in this case, and from the reduced form of $\partial_t\hat{\psi}''_2$, it can be seen that the $\ell=0$ and $\ell=1$ solutions will not yield radiative multipoles due to the vanishing of this Weyl scalar, i.e., the lowest multipole of gravitational radiation is the quadrupole ($\ell=2$) \cite{price1}. The relation shown in equation (\ref{dtpsi2}) was previously found in the case of perturbations of the Schwarzschild black hole in \cite{price2}. There, it was also shown that $\hat{\psi}''_2$ is invariant under infinitesimal null tetrad rotations and under gauge transformations as well, making this quantity measurable by any observer. Such properties are also valid for the $\hat{\psi}''_2$ of the gravitational perturbations discussed in this paper.

\section{V. The Morris-Thorne Wormholes}

In this section we will apply the master equation found in section III to the wormhole space-times introduced in \cite{mtwh}. The general line element in geometrized units ($G=c=1$) is the following,

\begin{equation}
ds^2=e^{2\Phi(r)}dt^2-\frac{dr^2}{1-b(r)/r}-r^2d\Omega^2,
\label{mt}
\end{equation}

where $\Phi(r)$ is known as the redshift function and $b(r)$ as the shape function. Both of these metric components fulfill certain conditions in order for the geometry of the space-time to be that of a wormhole. In particular, there exists a minimum radius $r=b_0>0$ such that $b(b_0)=b_0$. This value defines the throat of the wormhole, and hence, the domain of the radial coordinate is $r\in[b_0,\infty)$. It should be clarified that this coordinate decreases from positive infinity to $b_0$ as the throat is approached from one of the two universes it connects, and then increases back to infinity when emerging on the other universe. Additional requirements are that $1-b(r)/r\geq0$ for the shape function, along with $\Phi(r)$ being everywhere finite. This last condition on the redshift function is related to the non-existence of event horizons in the space-time so that hypothetical travelers may move from one universe to the other in both directions. If the wormhole is to be asymptotically flat, then the limits $\Phi(r)\rightarrow0$ and $b(r)/r\rightarrow0$ as $r\rightarrow\infty$ must also be imposed.

After this brief presentation on the geometrical features of the Morris-Thorne wormholes our intention next is to apply the reduced master equation (\ref{Qrad}) to them. Before that, a discussion of the properties of the gravitational source is in order. Following Morris and Thorne, we consider matter whose stress-energy tensor in an orthonormal frame is $$T_{\hat{\mu}\hat{\nu}}=diag\left[\rho,-\tau,p,p\right],$$ whose components are given a physical interpretation in which $\rho$ is the energy density, $\tau$ is the tension per unit area in the radial direction, and $p$ is the pressure in the lateral directions\footnote{For the sake of clarity we outline that, hereafter, the symbols $\rho$ and $\tau$ are no longer used for the spin coefficients of the Newman-Penrose formalism, but instead to represent now the mentioned physical quantities. The same will happen from equation (\ref{Ruv}) forward, where $\pi$ will denote the usual geometrical constant instead of the spin coefficient.}. All of these quantities are expressed as measured by a static observer and depend on the coordinate $r$. The energy density $\rho$ and the tension $\tau$ are given by (c.f. equations (17) and (18) in \cite{mtwh})

\begin{equation}
\rho=\frac{b'}{8\pi r^2}, \hspace{5mm} \tau=\frac{1}{8\pi r^2}\left(\frac{b}{r}-2(r-b)\Phi'\right).
\label{rhotau}
\end{equation} 

Morris and Thorne demonstrated that in order for the space-time to have the previously described geometric properties of a wormhole, the null energy condition must be violated at least near its throat, this is, $\tau(b_0)>\rho(b_0)$. The implication is that there exist observers that measure a negative energy density, this could include a static observer too. 

In the canonical $\{t,r,\theta,\varphi\}$ frame, the previous stress-energy tensor can be written as

\begin{equation}
T_{\mu\nu}=(\rho+p)u_\mu u_\nu-(\tau+p)v_\mu v_\nu-pg_{\mu\nu},
\label{Tuv}
\end{equation}

where $u^\mu$ is the 4-velocity of the matter in a co-moving frame and $v^\mu$ a unit space-like vector orthogonal to $u^\mu$ and pointing in the $x^1=r$ direction. If $\tau=-p$, the stress-energy tensor of a perfect fluid is recovered. From (\ref{Tuv}), the Ricci tensor is found to be 

\begin{equation}
\frac{1}{8\pi}R_{\mu\nu}=(\rho+p)u_\mu u_\nu-(\tau+p)v_\mu v_\nu-\frac{1}{2}(\rho+\tau)g_{\mu\nu}.
\label{Ruv}
\end{equation}

Comparing this expression to (\ref{Rmat}) and (\ref{Suv}), we have that $S_{\mu\nu}/8\pi=(\rho+p)u_\mu u_\nu-(\tau+p)v_\mu v_\nu$, and hence $c=0$. Thus, the master equation is valid for the Morris-Thorne wormholes, so long as the space-time is not electrically charged (see discussion after equation (\ref{RS})), and when the Ricci tensor can be casted in the form $(\ref{Ruv})$. The tortoise coordinate is then defined by

\begin{equation}
\frac{d}{dr_*}=\pm e^{\Phi}\sqrt{1-\frac{b}{r}}\frac{d}{dr}.
\nonumber
\end{equation}

In the coordinate transformation performed, one can always choose an adequate integration constant so that the throat of the wormhole is located at $r_*=0$. Moreover, the $r_*$ coordinate takes the positive sign for one side of the throat, and the negative sign for the other side. When $r\rightarrow\infty$, one has that $r_*\rightarrow\pm r$. The coordinate $r_*$ therefore takes values on the whole real line, i.e., $r_*\in(-\infty,\infty)$.

Inserting the metric components of line element (\ref{ds2}) in the general expression (\ref{V}) of the potential $V(r)$ we obtain,

\begin{equation}
V(r)=\frac{e^{2\Phi}}{r^2}\left[\ell(\ell+1)-\frac{5b}{2r}-r\Phi'\left(1-\frac{b}{r}\right)+\frac{b'}{2}\right].
\nonumber
\end{equation}

Using the relations (\ref{rhotau}) for the energy density and the tension, the potential can be rearranged as 

\begin{equation}
V(r)=\frac{e^{2\Phi}}{r^2}\left[\ell(\ell+1)-\frac{3b}{r}+4\pi r^2(\rho+\tau)\right].
\label{VMT}
\end{equation}

With the domain of the tortoise coordinate established and an explicit expression for the potential $V(r)$, the stability analysis consists now in studying the eigenvalue equation $\mathcal{H}R=\omega^2R$ with $\mathcal{H}=-d^2/dr_*^2+V(r)$. Specifically, if there exist eigenvalues which represent perturbations that grow without bound as $t\rightarrow\infty$, but are physically regular otherwise. Since the operator $\mathcal{H}$ is self-adjoint, the eigenvalues $\omega^2$ must be real. Hence, considering the time dependent part of the proposed ansatz, any instability will appear as a purely imaginary $\omega$, this is, as a negative eigenvalue.

A qualitative discussion of the eigenvalue spectrum of the operator $\mathcal{H}$ follows in a fairly simple manner based on the properties of the potential $V(r)$. If the potential is strictly positive there cannot exist negative eigenvalues (energy bound states) whose eigenfunctions are physically regular and thus, all of the vibrational modes of this class of wormholes are linearly stable, at least under odd-parity perturbations. In this case the eigenvalue spectrum is continuous. On the other hand, if $V(r)<0$ at some region of the space-time, it is possible that regular solutions with negative eigenvalues arise, leading to the instability of at least one of the vibrational modes of the wormhole.

By examining the individual terms that appear in the potential (\ref{VMT}) one can realize that a sufficiently negative energy density, which is possible due to the violation of the energy conditions, can make $V(r)<0$ for some coordinate values $r$. Thus, stability can be seen to strongly depend on the physical parameters $\rho$ and $\tau$. In what follows we will focus on a particular class of Morris-Thorne metrics, those for which $\rho+\tau=0$, as they will be proven to describe wormholes with no unstable modes of odd-parity gravitational perturbations.

The condition $\rho+\tau=0$ means physically that the energy density matches the radial pressure of the matter (the negative of the tension). This condition determines a constraint on the redshift and shape functions which can be expressed through equations (\ref{rhotau}), namely,

\begin{equation}
\frac{rb'(r)+b(r)}{2r}+\left(b(r)-r\right)\Phi'(r)=0.
\label{constmt}
\end{equation} 

We will now show that the class of Morris-Thorne metrics defined by (\ref{constmt}), satisfies the conditions that a wormhole must possess. The most compelling way to accomplish this is to rearrange the defining constraint of the class so that the shape function, without its first derivative, is in terms only of the redshift function. This will allow us to pick a suitable $\Phi(r)$, specifically an everywhere finite function, and find the corresponding expression for $b(r)$. Using the basic theory of first-order differential equations one can show that the desired relation between these functions is

\begin{equation}
b(r)=r-\frac{2e^{-2\Phi(r)}}{r}\left[F(r)+c_1\right],
\nonumber
\end{equation}

where $c_1$ is an integration constant and $$F(r)=\int re^{2\Phi(r)}dr.$$

The integration constant can be chosen so that the condition $b(b_0)=b_0$ on the minimum radius $r=b_0$ is fulfilled. Obtaining thus,

\begin{equation}
b(r)=r-\frac{2e^{-2\Phi(r)}}{r}\int_{b_0}^rr'e^{2\Phi(r')}dr'.
\label{b}
\end{equation} 

From (\ref{b}) and the fact that the integrand there is strictly positive in the domain of integration, it can be seen that the condition $1-b(r)/r\geq0$ is satisfied. This also implies that the vector $\partial/\partial r$ remains everywhere space-like. Furthermore, by examining the limit $r\rightarrow\infty$ for which $\Phi(r)\rightarrow0$, one can realize that $b(r)/r\rightarrow0$. Therefore, the wormhole fulfills the asymptotically flatness condition too. We have obtained thus a relation for the shape function in which, given an appropriate redshift function, the metrics of interest possess indeed the geometry of a wormhole. 

The potential of the Schr\"odinger operator for this restricted class of Morris-Thorne wormholes is now simply $$V(r)=\frac{e^{2\Phi}}{r^2}\left[\ell(\ell+1)-\frac{3b}{r}\right].$$

Recall that, for a vanishing value of the constant $c$ (as is the case), we concluded from the analysis in the previous section of the Weyl scalar $\psi_2$ that the lowest radiative multipole is the quadrupole. Then, it is readily seen that $V(r)\geq0$ for all $r\in[b_0,\infty)$, due to the $1-b(r)/r\geq0$ condition and to the fact that $\ell$ takes positive integer values starting from $\ell=2$. It can be shown that the asymptotic solution of the eigenfunctions is

\begin{equation}
R\sim r_*[h_\ell^{(1)}(r_*\omega_\pm)+h_\ell^{(2)}(r_*\omega_\pm)] \hspace{5mm} \text{as } r_*\rightarrow\pm\infty,
\nonumber
\end{equation}

where $h_\ell^{(1)}$ and $h_\ell^{(2)}$ denote the spherical Hankel functions of the first kind and of the second kind, respectively, and of order $\ell\geq2$. Given this behavior and since the potential is strictly positive, the eigenvalue spectrum of the operator $\mathcal{H}$ is positive and continuous. Also, by equation (\ref{dtpsi2}) and the peeling theorem, the eigenfunctions $R$ with positive eigenvalues ($\omega^2\geq0$) will describe physically regular perturbations. Thus, there are no linearly unstable vibrational modes generated by perturbations of odd-parity in this class of wormholes.

To finalize this section we provide some examples of this class of Morris-Thorne wormholes in table 1. They are easily obtained utilizing equation (\ref{b}) for the shape function. This process requires only of a well-behaved and bounded redshift function as input and so, can be used to yield as many space-times as functions that exist of this type. Note that the $\Phi(r)=0$ case reduces to the well-known Ellis wormhole\footnote{To obtain its more familiar line element, the transformation from the radial coordinate $r$ to $r_*=\pm\sqrt{r^2-b_0^2}$ is needed. In this case the $r_*$ coordinate is the proper radial distance.}, which additionally is a solution of the Einstein-scalar field equations with a negative sign. Unfortunately, since all of these wormholes belong to the family of Morris-Thorne metrics, they violate the energy conditions at least near their throats.

Interestingly enough, and though the $\ell=0$ modes do not yield gravitational radiation as a result of the perturbation, the potential $V(r)$ we deduce here reduces to that studied in \cite{phantom3,phantom4} for the Ellis metric when inserting the $\ell=0$ value. In those works the instability of that wormhole follows due to their corresponding potential being negative. This indicates that the angular dependance of the solution proposed here is crucial to deduce stability, at least for the odd-parity case. Of course, the reason why we obtain a different result lies in the type of perturbation we have analyzed during this work. 

\begin{table}
	\caption{Metric components of a few examples from the class of Morris-Thorne wormholes studied in section V.}
	\begin{center}
		\begin{tabular}{ |c|c| }
		\hline
		$e^{2\Phi(r)}$ & $1-b(r)/r$ \\ \hline \hline
		$1$ & $1-b_0^2/r^2$ \\ \hline  
		$1+e^{-(r/b_0)^2}$ & $e^{-2\Phi(r)}[1-b_0^2(e^{2\Phi(r)}-e^{-1})/r^2]$ \\ \hline 
		$1+b_0^2/(x^2+b_0^2)$ & $e^{-2\Phi(r)}\left[1+b_0^2(\ln\left[1/2+r^2/2b_0^2\right]-1)/r^2\right]$ \\ \hline
		$1/2+\arctan(r/b_0-1)/\pi$ & $e^{-2\Phi(r)}\left[e^{2\Phi(r)}-b_0/\pi r-b_0^2\left(\pi/2-1+\ln[1+(1-r/b_0)^2]\right)/\pi r^2\right]$	\\ \hline
		\end{tabular}
		\end{center}
	\label{tab:ejemplos}
\end{table}

\section{VI. Odd-Parity Perturbations in a Phantom Scalar Field Wormhole}

In section III we mentioned that the master equation derived there is valid for solutions of the Einstein-scalar field equations. In fact, one of the examples of Morris-Thorne wormholes shown in table 1 is indeed a solution of this type, namely the Ellis metric. In what follows we will examine the perturbation equation of one last example of a wormhole supported by a phantom scalar field, i.e., a solution to $R_{\mu\nu}=-\nabla_\mu\phi\nabla_\nu\phi$. This space-time was found by Ellis and Bronnikov \cite{ellis,Bronnikov}, a generalization to rotating scalar field wormholes was later introduced in \cite{Matos}. Here, we will focus on its static version since our master equation can only be applied to that reduced form of the metric. Its line element in Boyer-Lindquist coordinates is

\begin{equation}
ds^2=fdt^2-\frac{1}{f}\left[dr^2+(r^2-2rr_1+r_0^2)d\Omega^2\right],
\nonumber
\end{equation}

with $f=e^{-\phi_0(\lambda-\pi/2)}$ and $\lambda=\arctan\left[(r-r_1)/\sqrt{r_0^2-r_1^2}\right]$. In this coordinate system we have for the Boyer-Lindquist radius that $-\infty<r<\infty$, covering both universes this way. The quantities $r_0$ and $r_1$ are constant parameters whose units are that of length, and for which $r_0^2>r_1^2$. The scalar field is given by $\phi=\sqrt{2+\phi_0^2/2}(\lambda-\pi/2)$, being $\phi_0$ a constant without units. In this wormhole the throat joins two asymptotically flat sides, nevertheless, these sides are not symmetrical. This can be seen when taking the asymptotic limits of the $f$ function, $$\lim_{r\rightarrow\infty}f=1, \hspace{4mm} \lim_{r\rightarrow-\infty}f=e^{\phi_0\pi}.$$ 

By rescaling the $t$ and $r$ coordinates to $t_-=e^{\phi_0\pi/2}t$ and $r_-=e^{-\phi_0\pi/2}r$, it can be realized that indeed the other side of the throat is asymptotically flat as well. The wormhole becomes symmetric only if $\phi_0=0$, in which case, the line element reduces to that of the Ellis metric. It will be convenient to replace the coordinate $r$ with $x=(r-r_1)/L$, where $L^2=r_0^2-r_1^2$. Thus,

\begin{equation}
ds^2=fdt^2-\frac{L^2}{f}\left[dx^2+(x^2+1)d\Omega^2\right],
\label{ds2sfwh}
\end{equation}

and $\lambda=\arctan x$. In these coordinates the throat of the wormhole is located at $x=0$, while the upper and lower universes are described by $x>0$ and $x<0$, respectively.

To obtain the equation that governs the odd-parity gravitational perturbations of this space-time we proceed with the same scheme as in the previous section. The tortoise coordinate is given by 

\begin{equation}
\frac{d}{dx_*}=\frac{f}{L}\frac{d}{dx}.
\nonumber
\end{equation}

Since $f$ is regular for all $x\in\mathbb{R}$ and because of the asymptotic form of said function at both infinities, the new coordinate ranges over the values $-\infty<x_*<\infty$. A suitable integration constant can also be picked so that the throat is described by $x_*=0$. Assuming a similar ansatz as the one used throughout this paper, $Q=e^{i\omega t}X(x)\sin\theta dP_\ell(\cos\theta)/d\theta$, and substituting the metric functions of the phantom wormhole in (\ref{V}), we have that

\begin{equation}
\frac{d^2X}{dx_*^2}-\left(V(x)-\omega^2\right)X=0,
\label{Qsfwh}
\end{equation}

where now

\begin{equation}
V(x)=\frac{f^2}{L^2(x^2+1)}\left[\ell(\ell+1)+\frac{3}{x^2+1}\left(\phi_0x+\frac{\phi_0^2}{4}-1\right)\right].
\nonumber
\end{equation}

In this case equation (\ref{Qsfwh}) defines an eigenvalue problem for the operator $\mathcal{H}=-d^2/dx_*^2+V(x)$. This potential coincides with that found in \cite{YanChew} for the case of axial perturbations. Its properties are the same as those of the previous examples in section V. Additionally, it can be easily verified that the second term that appears inside brackets in the expression of $V(x)$ has a global minimum $u_{min}=-3$ at the coordinate value $x=-\phi_0/2$. Hence, appealing to the fact that the $\ell=2$ vibrational modes are the lowest possible, the potential $V(x)$ is strictly positive for all $x\in\mathbb{R}$. By the same arguments as those mentioned for the former class of Morris-Thorne wormholes, we can conclude that this scalar field wormhole is stable when its metric is perturbed by a small term of odd-parity.

As mentioned before, and just like in the class of Morris-Thorne metrics previously discussed, the Ellis space-time is again a particular case of this phantom scalar field wormhole when the parameter $\phi_0$ vanishes\footnote{For this particular case the relation between the proper radial length $r_*$ and the coordinates $x=x_*$ is $Lx=r_*$, with $L=b_0$.}. The wormhole presented here also violates the energy conditions as a result of it being a solution of the Einstein-scalar field equations with a negative sign.

\section{Conclusions and Some Additional Comments}

We have utilized the Newman-Penrose formalism to obtain a so-called master equation that describes the linear behavior of gravitational perturbations in stationary and spherically symmetric space-times. The perturbations were assumed to be of odd-parity in the Regge-Wheeler gauge. This framework allowed us to write the derived master equation in a compact (and may we dare say elegant) manner through the use of the spin coefficients and operators that characterize the formalism. Our master equation is not applicable, though, to the whole generality of space-times with spherical symmetry, this is due to a constraint on certain components of the Ricci tensor that has to be obeyed. Despite this, we showed that it is well-suited to analyze some interesting examples of metrics that describe wormholes, for instance, the solutions of the Einstein-scalar field equations. Other space-times that were found to be within the range of validity of our master equation belong to the family of Morris-Thorne wormholes. We focused on those metrics whose gravitational source have the particular property that its energy density is equal to its radial pressure. After applying the aforementioned master equation to them, we found that there are no unstable modes of vibration due to odd-parity perturbations. The explicit metric components of some of this type of space-times were presented too. Finally, we gave one last example of a static scalar field wormhole that, according to the properties of its corresponding master equation, is not unstable against the perturbations here studied.

It should be borne in mind that, while our results indicate stability for some wormholes, it is only with respect to perturbations of odd-parity. Future developments of this work include the study of their even-parity counterparts within the Newman-Penrose formalism. However, the complexity of the calculations involved for this purpose increases compared to the odd case. Another interesting aspect to determine is the possibility to generalize the scheme presented here for gravitational perturbations in the context of the tetrad formalism to axially symmetric space-times. This in turn implies a generalization of the Regge-Wheeler gauge to this kind of metrics. Yet again, the whole process may require of lengthy calculations that hopefully are still manageable from an analytical approach.

\section{Appendix A}
\renewcommand{\theequation}{A.\arabic{equation}}
\setcounter{equation}{0}

Here we show all the relevant quantities of the Newman-Penrose formalism calculated for metric (\ref{ds2}) with background tetrad (\ref{tetrada}) and perturbation matrix (\ref{sigma}). We will reference the equations of the NP paper from which our results are derived. To simplify notation, the use of the tilde for background quantities will be dropped and the hat will be kept for the perturbation terms. Thus, any quantity or operator without a hat should be understood to be of the background space-time, except for the perturbations functions $f_0$ and $f_1$.

From (NP 4.1a) and equation (\ref{zabc2}) of our text, the perturbation term of the spin coefficients is given by

\begin{align}
\hat{\kappa}&=\hat{\mathcal{Z}}_{020}=Df_1-\frac{1}{2}\delta f_0, & \hat{\pi}&=-\hat{\mathcal{Z}}_{031}=0, \nonumber\\
\hat{\nu}&=-\hat{\mathcal{Z}}_{131}=\Delta f_1-\frac{1}{2}\delta^*f_0, & \hat{\tau}&=\hat{\mathcal{Z}}_{120}=0, \nonumber\\
\hat{\rho}&=\hat{\mathcal{Z}}_{320}=(-\delta_-+\kappa_+-\pi_+)f_1, & \hat{\lambda}&=-\hat{\mathcal{Z}}_{331}=0, \nonumber\\
\hat{\mu}&=-\hat{\mathcal{Z}}_{231}=(\delta_--\kappa_++\pi_+)f_1, & \hat{\sigma}&=\hat{\mathcal{Z}}_{220}=0, \nonumber\\
\hat{\alpha}&=\frac{1}{2}(\hat{\mathcal{Z}}_{310}-\hat{\mathcal{Z}}_{332})=\frac{1}{2}(\nu f_0-Df_1), & \hat{\beta}&=\frac{1}{2}(\hat{\mathcal{Z}}_{210}-\hat{\mathcal{Z}}_{232})=-\frac{1}{2}(\kappa f_0-Df_1), \nonumber\\
\hat{\varepsilon}&=\frac{1}{2}(\hat{\mathcal{Z}}_{010}-\hat{\mathcal{Z}}_{032})=\frac{1}{2}\left((-\delta_-+\kappa_+-\pi_+)f_1-\Delta f_0\right), \nonumber\\ 
\hat{\gamma}&=\frac{1}{2}(\hat{\mathcal{Z}}_{110}-\hat{\mathcal{Z}}_{132})=\frac{1}{2}\left((\delta_--\kappa_++\pi_+)f_1-Df_0\right), 
\label{spin}
\end{align} 

with the definitions $\delta_{\pm}=(\delta\pm\delta^*)/2$, $\kappa_{\pm}=(\kappa\pm\nu)/2$, and $\pi_{\pm}=(\pi\pm\tau)/2$. In terms of the metric components these newly defined coefficients and operators take the explicit form

\begin{eqnarray}
\kappa_+=-\pi_+&=&\frac{i\cot\theta}{2\sqrt{2g_2}}, \hspace{4mm} \pi_-+\kappa_-=-2\alpha=\frac{g'_2}{2g_2\sqrt{2g_1}}, \hspace{4mm} \pi_--\kappa_-=\frac{g'_0}{2g_0\sqrt{2g_1}}, \nonumber\\
\delta_+&=&\frac{1}{\sqrt{2g_1}}\frac{\partial}{\partial r}, \hspace{10mm} \delta_-=\frac{i}{\sqrt{2g_2}}\frac{\partial}{\partial\theta}, \hspace{10mm} D=\frac{1}{\sqrt{2}}\left(\frac{1}{\sqrt{g_0}}\frac{\partial}{\partial t}+\frac{1}{\sqrt{g_2}\sin\theta}\frac{\partial}{\partial\varphi}\right),
\label{spinoppm}
\end{eqnarray}

where a prime in this set of equations denotes derivation with respect to the radial coordinate $r$. Note that $\delta_+^*=\delta_+$, $\delta_-^*=-\delta_-$, and that $\kappa_-,\pi_-\in\mathbb{R}$ while $\kappa_+$, $\pi_+$ are purely imaginary. With this notation, identity (\ref{id2}) can be expressed as

\begin{equation}
\delta_+\kappa_+=2\alpha\kappa_+.
\label{id2pm}
\end{equation}

The linearized perturbed components of the Ricci tensor in tetrad form, sometimes called the Ricci identities, can be computed by the Newman-Penrose equations (NP 4.2). Thereby, we obtain

\begin{align}
D\hat{\rho}-\delta^*\hat{\kappa}&=-\kappa^*\hat{\tau}-\hat{\kappa}^*\tau-\kappa(3\hat{\alpha}+\hat{\beta}^*-\hat{\pi})-\hat{\kappa}(3\alpha+\beta^*-\pi)+\hat{\Phi}_{00},
\tag{NP 4.2a}\\
D\hat{\sigma}-\delta\hat{\kappa}&=-(\tau-\pi^*+\alpha^*+3\beta)\hat{\kappa}-(\hat{\tau}-\hat{\pi}^*+\hat{\alpha}^*+3\hat{\beta})\kappa+\hat{\psi}_0,
\tag{NP 4.2b}\\
D\hat{\alpha}-\delta^*\hat{\varepsilon}&=(\hat{\rho}+\hat{\varepsilon}^*-2\hat{\varepsilon})\alpha+\beta\hat{\sigma}^*-\beta^*\hat{\varepsilon}-\kappa\hat{\lambda}-\kappa^*\hat{\gamma}+(\hat{\varepsilon}+\hat{\rho})\pi+\hat{\Phi}_{10},
\tag{NP 4.2d}\\
D\hat{\gamma}-\Delta\hat{\varepsilon}&=(\hat{\tau}+\hat{\pi}^*)\alpha+(\hat{\tau}^*+\hat{\pi})\beta+\tau\hat{\pi}+\hat{\tau}\pi-\nu\hat{\kappa}-\hat{\nu}\kappa+\hat{\Psi}_2-\hat{\Lambda}+\hat{\Phi}_{11},
\tag{NP 4.2f}\\
D\hat{\lambda}-\delta^*\hat{\pi}&=2\pi\hat{\pi}+(\alpha-\beta^*)\hat{\pi}+(\hat{\alpha}-\hat{\beta}^*)\pi-\nu\hat{\kappa}^*-\hat{\nu}\kappa^*+\hat{\Phi}_{20},
\tag{NP 4.2g}\\
D\hat{\mu}-\delta\hat{\pi}&=\pi\hat{\pi}^*+\hat{\pi}\pi^*-\pi(\hat{\alpha}^*-\hat{\beta})-\hat{\pi}(\alpha^*-\beta)-\nu\hat{\kappa}-\hat{\nu}\kappa+\hat{\Psi}_2+2\hat{\Lambda},
\tag{NP 4.2h}\\
\Delta\hat{\lambda}-\delta^*\hat{\nu}&=(3\alpha+\beta^*+\pi-\tau^*)\hat{\nu}+(3\hat{\alpha}+\hat{\beta}^*+\hat{\pi}-\hat{\tau}^*)\nu+\hat{\psi}_4,
\tag{NP 4.2j}\\
\delta\hat{\alpha}-\delta^*\hat{\beta}&=\alpha\hat{\alpha}^*+\hat{\alpha}\alpha^*+\beta\hat{\beta}^*+\hat{\beta}\beta^*-2\alpha\hat{\beta}-2\hat{\alpha}\beta-\hat{\Psi}_2+\hat{\Lambda}+\hat{\Phi}_{11},
\tag{NP 4.2l}\\
\delta\hat{\nu}-\Delta\hat{\mu}&=-\nu^*\hat{\pi}-\hat{\nu}^*\pi+(\tau-3\beta-\alpha^*)\hat{\nu}+(\hat{\tau}-3\hat{\beta}-\hat{\alpha}^*)\nu+\hat{\Phi}_{22},
\tag{NP 4.2n}\\
\delta\hat{\gamma}-\Delta\hat{\beta}&=(\tau-\beta-\alpha^*)\hat{\gamma}+\hat{\mu}\tau-\hat{\sigma}\nu-\hat{\varepsilon}\nu^*-\beta(\hat{\gamma}-\hat{\gamma}^*-\hat{\mu})+\alpha\hat{\lambda}^*+\hat{\Phi}_{12},
\tag{NP 4.2o}\\
\delta\hat{\tau}-\Delta\hat{\sigma}&=(\tau+\beta-\alpha^*)\hat{\tau}+(\hat{\tau}+\hat{\beta}-\hat{\alpha}^*)\tau-\kappa\hat{\nu}^*-\hat{\kappa}\nu^*+\hat{\Phi}_{02},
\tag{NP 4.2p}
\end{align}

where we have taken advantage of the property $\hat{D}_m\phi=0$ that our particular choice of tetrad gives us for arbitrary background scalars $\phi$. We have also omitted the background terms that should appear on both sides of these equations since they cancel each other out.

The commutators (NP 4.4) of the background differential operators of the formalism are

\begin{equation}
[\Delta,D]=0, \hspace{4mm} [\delta,D]=-\pi^*D+\kappa\Delta, \hspace{4mm} [\delta,\Delta]=-\nu^*D+\tau\Delta, \hspace{4mm} [\delta^*,\delta]=2\alpha(\delta-\delta^*),
\label{conm}
\end{equation}

which can be utilized to derived the commutation relations for our previously introduced operators $\delta_\pm$,

\begin{equation}
[\delta_{\pm},D]=\mp\pi_{\mp}D+\kappa_{\mp}\Delta, \hspace{4mm} [\delta_{\pm},\Delta]=\kappa_{\mp}D\mp\pi_{\mp}\Delta, \hspace{4mm} [\delta_-,\delta_+]=-2\alpha\delta_-.
\label{conmpm}
\end{equation}

When applying these commutators to $\varphi$-independent scalar quantities $\phi$, as will always be the case in this work, there is a further simplification $[\delta_-,D]\phi=[\delta_-,\Delta]\phi=0$, since $D\phi=\Delta\phi$ and $\kappa_++\pi_+=0$. 

After some considerable algebraic steps, reduced equations for the linearized Ricci identities can be obtained by inserting the perturbed spin coefficients (\ref{spin}) into the (NP 4.2) equations presented above, along with the further aid of the commutators in (\ref{conmpm}) and the spin coefficient properties (\ref{id}, \ref{PhiA}). Doing so yields

\begin{align}
\hat{\Phi}_{00}=-\hat{\Phi}_{22}=&\frac{1}{2}\left[(\delta_++\kappa_-+3\pi_-)\delta_+-(\delta_-+2\kappa_+)\delta_-+4(\kappa_+^2-\kappa_-^2)\right]f_0 \nonumber\\
&-(\delta_+-6\alpha)Df_1, \nonumber\\
\hat{\Phi}_{12}=\hat{\Phi}_{21}^*=-\hat{\Phi}_{01}=-\hat{\Phi}_{10}^*=&\frac{1}{2}\left[D^2+(\delta_++\delta_-+\kappa_-+3\pi_-+4\kappa_+)(\delta_--2\kappa_+)\right]f_1 \nonumber\\
&-\frac{1}{4}(\delta_++\delta_-+\pi_--3\kappa_--2\kappa_+)Df_0, \nonumber\\
\hat{\Phi}_{11}=\hat{\Phi}_{20}=\hat{\Phi}_{02}^*=\hat{\Lambda}=&0.
\label{PhiB}
\end{align}

For the Weyl scalars of interest we obtain

\begin{eqnarray}
\hat{\psi}_0=-\hat{\psi}_4^*&=&\frac{1}{2}\left(\delta_++\delta_--\kappa_-+\pi_-+2\kappa_+\right)\left[(\delta_++\delta_-)f_0-2Df_1\right]-2(\kappa_++\kappa_-)\left[Df_1+(\kappa_++\kappa_-)f_0\right], \nonumber\\
\hat{\psi}_2&=&\delta_-Df_1+(\kappa_-\delta_--\kappa_+\delta_+)f_0.
\label{psi}
\end{eqnarray}

\section{Appendix B}
\renewcommand{\theequation}{B.\arabic{equation}}
\setcounter{equation}{0}

A more detailed proof of the consistency condition (\ref{R03c}) of the linearized Einstein field equations is presented in this appendix. 

When applying the operator $D$ to it, and using the commutators (\ref{conm}), the component $\hat{\mathcal{R}}_{03}$ of the system (\ref{R3mat}) reduces to

\begin{align}
0=&2(\delta_+-4\alpha+2\pi_-)D^2f_1 \nonumber\\
&+\left[(\delta_-+2\kappa_+)\delta_--(\delta_++4\pi_-)(\delta_++\pi_--\kappa_-)+4(\kappa_-^2-\kappa_+^2)+2(3\Lambda-\Phi_{11})-S\right]Df_0.
\label{DR03B}
\end{align}

Expression (\ref{Df0}) for the perturbation function $f_0$ can now be substituted in (\ref{DR03B}). The resulting terms can be rearranged as 

\begin{align}
0=&(\delta_+-4\alpha+2\pi_-)\left([D^2+(\delta_-+4\kappa_+)(\delta_--2\kappa_+)-(\delta_+-3\kappa_-+\pi_-)(\delta_++2\pi_-)]f_1+2f_1[\delta_-+4\kappa_+]\kappa_+\right) \nonumber\\
&-Df_0\left[(\delta_++\kappa_-+3\pi_-)\kappa_-+2\kappa_+^2-3\Lambda+\Phi_{11}+S/2\right].
\nonumber
\end{align}

Careful attention must be paid on the order in which the operators are being applied. The previous equation can be simplified by using (\ref{f1}) and defining the quantities $\mathcal{A}=2(\delta_-+4\kappa_+)\kappa_+-2(3\Lambda-\Phi_{11}+\Phi_{00})+S$, as well as $\mathcal{B}=2(\delta_++\kappa_-+3\pi_-)\kappa_-+4\kappa_+^2-2(3\Lambda+\Phi_{11})+S$. Despite the appearance of differential operators in these quantities, $\mathcal{A}$ and $\mathcal{B}$ should not be understood as such. They are merely scalar quantities, the operators $\delta_\pm$ in them are to be applied only to the spin coefficients $\kappa_{\pm}$. Hence, we can write

\begin{equation}
\left[(\delta_+-4\alpha+2\pi_-)\mathcal{A}-\mathcal{B}(\delta_++2\pi_-)\right]f_1=0.
\label{DR03B2}
\end{equation}

Expanding the first term of (\ref{DR03B2}) results in

\begin{equation}
f_1\left[(\delta_+-4\alpha+2\pi_-)\mathcal{A}-2\pi_-\mathcal{B}\right]+2(\mathcal{A}-\mathcal{B})\delta_+f_1=0.
\label{DR03B3}
\end{equation}

Using the background Ricci identities of the Newman-Penrose formalism (NP 4.2a) and (NP 4.2b), it can be proven that $\mathcal{A}=\mathcal{B}=4\kappa_+^2-(\psi_0+\psi_0^*)/2-\Phi_{00}-2(3\Lambda-\Phi_{11})+S$. Another helpful identity, consequence of (\ref{id2pm}) and (\ref{conmpm}), is $(\delta_+-4\alpha)(\delta_-+4\kappa_+)\kappa_+=0$. Equation (\ref{DR03B3}) thereby simplifies to

\begin{equation}
f_1(\delta_+-4\alpha)(S-2\Lambda_s)=0,
\nonumber
\end{equation}

with $\Lambda_s=3\Lambda-\Phi_{11}+\Phi_{00}$. By considering the explicit form of the spin coefficients and operators shown in (\ref{spinoppm}), and since $f_1$ cannot vanish, this past condition can be rewritten as $$\frac{1}{g_2\sqrt{2g_1}}\frac{d}{dr}\left[g_2(S-2\Lambda_s)\right]=0.$$

If this equation is true everywhere in space-time the implication is that

\begin{equation}
S-2\Lambda_s=c/g_2,
\label{const}
\end{equation}

where $c$ is an integration constant. Furthermore, using the (NP 4.3b) equations, we have that $R_{\mu\nu}l^\mu n^\nu=2(3\Lambda-\Phi_{11})$ and $R_{\mu\nu}l^\mu l^\nu=-2\Phi_{00}$. From the particular tetrad considered here and the expression (\ref{Suv}) for $S_{\mu\nu}$, it can be seen that 

\begin{equation}
S-2\Lambda_s=S_{22}/g_2.
\label{const2}
\end{equation}

Comparing (\ref{const}) with (\ref{const2}), the result that was anticipated in section III of the main text is obtained, i.e., $S_{22}=c$. \\


 \textbf{Acknowledgments.} This work was partially supported by CONACyT M\'exico under grants CB-2011 No. 166212, CB-2014-01 No. 240512, Project
No. 269652, Fronteras Project 281, and grant No. I0101/131/07 C-234/07 of the Instituto Avanzado de Cosmolog\'ia (IAC) collaboration (http://www.iac.edu.mx/). J.C.A. acknowledges financial support from CONACyT doctoral fellowships too. \\

\end{document}